\documentclass[lettersize,journal]{IEEEtran}
\usepackage{amsmath,amsfonts}
\usepackage{algorithmic}
\usepackage{algorithm}
\usepackage{array}
\usepackage[caption=false,font=normalsize,labelfont=sf,textfont=sf]{subfig}
\usepackage{textcomp}
\usepackage{stfloats}
\usepackage{url}
\usepackage{verbatim}
\usepackage{graphicx}
\usepackage{cite}
\usepackage{amssymb}
\usepackage{url} 
\usepackage{mathtools}
\usepackage{bm}
\usepackage{mathtools}
\usepackage{booktabs}
\usepackage{graphics}
\usepackage{wrapfig}

\usepackage[export]{adjustbox}
\usepackage{algorithm}
\usepackage{algorithmic}
\usepackage{multirow}
\usepackage{float}
\usepackage{soul}
\usepackage{comment}

\usepackage[table]{xcolor}
\usepackage{xcolor}
\begin{document}

\title{SpeechPrompt: Prompting Speech Language Models for Speech Processing Tasks}

\author{Kai-Wei Chang, Haibin Wu, Yu-Kai Wang, Yuan-Kuei Wu, Hua Shen, Wei-Cheng Tseng, Iu-thing Kang, Shang-Wen Li, Hung-yi Lee}


\newcommand{\proofread}[1]{{\small\textcolor{orange}{\bf [#1]}}}
\newcommand{\units}[1]{{\small\textcolor{green}{\bf [#1]}}}

\maketitle
\begin{abstract}
Prompting has become a practical method for utilizing pre-trained language models (LMs). This approach offers several advantages. It allows an LM to adapt to new tasks with minimal training and parameter updates, thus achieving efficiency in both storage and computation. Additionally, prompting modifies only the LM's inputs and harnesses the generative capabilities of language models to address various downstream tasks in a unified manner. This significantly reduces the need for human labor in designing task-specific models. These advantages become even more evident as the number of tasks served by the LM scales up. Motivated by the strengths of prompting, we are the first to explore the potential of prompting speech LMs in the domain of speech processing. Recently, there has been a growing interest in converting speech into discrete units for language modeling. Our pioneer research demonstrates that these quantized speech units are highly versatile within our unified prompting framework. Not only can they serve as class labels, but they also contain rich phonetic information that can be re-synthesized back into speech signals for speech generation tasks. Specifically, we reformulate speech processing tasks into speech-to-unit generation tasks. As a result, we can seamlessly integrate tasks such as speech classification, sequence generation, and speech generation within a single, unified prompting framework. The experiment results show that the prompting method can achieve competitive performance compared to the strong fine-tuning method based on self-supervised learning models with a similar number of trainable parameters. The prompting method also shows promising results in the few-shot setting. Moreover, with the advanced speech LMs coming into the stage, the proposed prompting framework attains great potential.

\end{abstract}
\begin{IEEEkeywords}
Prompting, speech language model, self-supervised learning, representation learning
\end{IEEEkeywords}


\section{Introduction}
Recently, self-supervised representation learning has become an essential component in the speech processing field~\cite{mohamed2022self}.
The speech \emph{representation model} is trained on a large-scale unlabeled corpus in a self-supervised learning (SSL) manner. 
The learned representation has been demonstrated to be informative and can benefit a wide range of speech processing tasks~\cite{DBLP:conf/interspeech/YangCCLLLLSCLHT21, DBLP:conf/interspeech/EvainNLBMATTDPA21, DBLP:conf/acl/TsaiCHHLYDLLSCH22}.

\begin{figure}[t]
    \centering
    \includegraphics[width=0.48\textwidth]{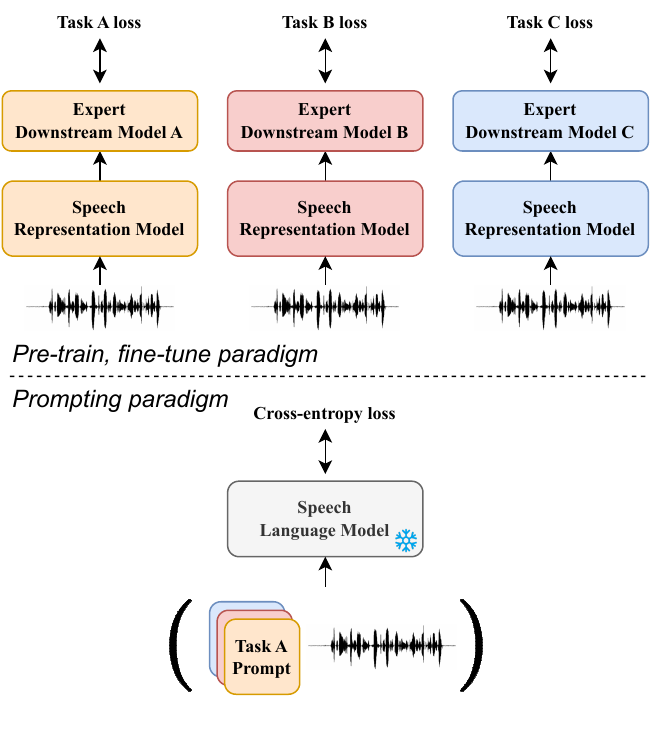}
    \caption{Comparison of the ``pre-train, fine-tune'' paradigm with the prompting paradigm. The ``pre-train, fine-tune'' paradigm involves designing task-specific downstream models and loss functions by human experts, with distinct models trained for each task. In contrast, the prompting paradigm handles all downstream tasks in a unified manner, where only the prompt varies for each task, while the language model remains fixed.}
    \label{fig:paradigm}
\end{figure}

When leveraging these speech representation models for a downstream task of interest, a typical approach is to follow the \textbf{``pre-train, fine-tune''} paradigm~\cite{mohamed2022self, liu2023pre}. 
Under this paradigm, the representation models serve as feature extractors. 
The models encode speech into informative representations, which are subsequently fed into a task-specific model.
This model, referred to as the \emph{expert downstream model}, specializes in solving a specific speech processing task.
While fine-tuning often yields optimal performance, this paradigm, as depicted in Fig.~\ref{fig:paradigm}, requires delicately designing a task-specific downstream model and loss function for each task.
This complexity significantly causes an increasing burden of human labor.
Furthermore, the requirement to train the expert downstream model alongside the optionally fine-tuned speech representation model leads to substantial computational and storage demands. This is especially challenging as the number of downstream tasks grows due to the necessity to store separate model parameters for each task.

On the other hand, researchers have explored the \textbf{``prompting paradigm''} \cite{liu2023pre} as an alternative method to leverage pre-trained language models (LMs) to solve downstream tasks in an efficient manner. 
Originating from the Natural Language Processing (NLP) field, prompting refers to the technique that finds a task-specific template or instruction, which is called \textbf{prompt}, to steer a pre-trained LM without modifying its architecture and parameters.
For each specific task, these templates can be hand-crafted or identified through a search process and are composed of the model's vocabulary, known as \emph{hard prompts}~\cite{DBLP:conf/emnlp/ShinRLWS20, DBLP:journals/jmlr/RaffelSRLNMZLL20}.
For instance, in sentiment classification, an input sentence $\langle S \rangle$ can be fit into a template: ``$\langle S \rangle$. It was \_\_.'' and then fed into a pre-trained LM.
The LM's output (e.g., ``great'', ``terrible'') is then transformed into sentiment classes (positive, negative) by a \emph{verbalizer}~\cite{DBLP:conf/naacl/SchickS21, DBLP:conf/eacl/SchickS21}, which is often a hand-crafted or a searched mapping function~\cite{hu2022knowledgeable}, enabling us to determine the sentiment of $\langle S \rangle$.
Alternatively, prompts are not necessarily to be human-readable. Researchers have proposed a prompting method known as \textbf{prompt tuning}, which involves learning continuous prompts~\cite{liu2023pre, ding2022delta, li2021prefix, liu2023gpt, lester2021power} within the model's embedding space. These prompt vectors, also called \emph{soft prompts}, are trainable and have shown to be effective and efficient for leveraging pre-trained models, with applications extending beyond the NLP field. For example, prompt tuning has been applied to computer vision~\cite{jia2022visual} and speech processing~\cite{chang22e_interspeech}. 

The prompting paradigm presents multiple advantages compared to the traditional ``pre-train, fine-tune'' paradigm: 

(1) \textbf{Training Efficiency:} Only prompt vectors require updating, offering better computational efficiency than the full model and downstream head training in the typical fine-tuning paradigm. 
Moreover, reformulating downstream tasks into a unified sequence generation task eliminates the need for \emph{Expert Model Engineering} (i.e., designing specialized downstream models for each task) and \emph{Objective Engineering} (i.e., designing loss functions for each downstream task).

(2) \textbf{Inference Uniformity:} With prompting, the LM remains fixed, enabling a uniform forward process for diverse tasks.
The task specificity is driven by the input prompts, facilitating \emph{in-batch tasking}~\cite{lester2021power, ding2022delta}, the concurrent handling of multiple tasks within a single batch. 

(3) \textbf{Deployment Scalability:} Recently, language models are increasingly deployed as services. The low computational and storage demands of prompting offer significant advantages. This is because the LM does not require retraining when serving a user's own dataset and task; instead, only task-specific prompts containing a small set of parameters need to be identified. As the number of tasks or users grows, the scalability and efficiency of prompting become even more beneficial~\cite{sun2022black}.
The advantages of both the prompting paradigm and the ``pre-train, fine-tune'' paradigm are illustrated in Table~\ref{tab:paradigm_comparison}.
\begin{table}[t]
\centering
\caption{Comparative analysis of prompting and ``pre-train, fine-tune'' paradigms across various criteria. Symbols used: \checkmark indicates a relative advantage. $\triangle$ denotes comparable performance. \(\times\) indicates no significant advantage.}
\begin{tabular}{lcc}
\toprule
\textbf{Criterion}                            & \textbf{Prompting} & \textbf{Pre-train Fine-tune} \\ \toprule
Objective Engineering                               & \(\times\)                  & \checkmark                           \\ 
Expert Model Engineering                       & \(\times\)                  & \checkmark                           \\ 
Task-Specific Performance                       & $\triangle$ & 	 \checkmark   \\
Low-resource Performance                       & 	\checkmark & $\triangle$ \\
Storage Efficiency                           & \checkmark                  & \(\times\)                           \\ 
Computation Efficiency                       & \checkmark                  & \(\times\)                           \\ 
Deployment Efficiency & \checkmark                  & \(\times\)                           \\ \toprule
\end{tabular}
\label{tab:paradigm_comparison}
\vspace{-0.3cm}
\end{table}

In the table, we also compare the task-specific performance and low-resource performance. The “pre-train, fine-tune” paradigm delicately performs expert model and objective engineering. Therefore, it usually shows advantages when one wants to achieve better performance for a specific downstream task. On the other hand, prompting utilizes the prior knowledge of the LM, therefore usually achieving better performance in low-resource settings, such as the few-shot learning scenario.

This paper focuses on prompting the \emph{textless speech language models}~\cite{lakhotia2021generative, kharitonov2021text, popuri22_interspeech}. 
These models are a class of generative LMs that are trained on discrete speech units obtained by quantizing the SSL speech representations~\cite{DBLP:conf/interspeech/PolyakACKLHMD21}.
Discrete speech units have gained researchers' attention because they offer several advantages: (1) Discrete units require less storage space and transmission bitrate compared to raw waveforms~\cite{chang23b_interspeech}. (2) Discrete units contain essential acoustic and linguistic information while minimizing speaker-specific information~\cite{DBLP:conf/interspeech/PolyakACKLHMD21}, which is useful for scenarios where privacy is a major concern~\cite{nautsch2019preserving, chang23b_interspeech}.
Mirroring the text LMs in the NLP field, these textless speech LMs adopt discrete units as their vocabulary and undergo pre-training through tasks like next token prediction~\cite{radford2018improving} and the denoising sequence-to-sequence~\cite{DBLP:conf/acl/LewisLGGMLSZ20} task.  
Thanks to these speech LMs, several works have demonstrated promising results in challenging speech processing tasks, including speech continuation~\cite{lakhotia2021generative} and speech-to-speech translation~\cite{popuri22_interspeech}— tasks that are hard to achieve with the traditional ``pre-train, fine-tune'' paradigm.
The textless property is particularly compelling since many languages worldwide lack substantial text resources~\cite{dunbar2022self}. 
These languages may either have no written form or lack a standardized written format. 
By directly modeling the phonetic and acoustic patterns, we can not only bypass the constraints and potential biases of written languages but also reduce the need for paired speech-text data, which is often costly to get.

Furthermore, the ability of these discrete units to encapsulate both acoustic and linguistic~\cite{DBLP:conf/interspeech/PolyakACKLHMD21, lakhotia2021generative} information without text supervision has opened up new opportunities for prompting the speech LM for a variety of speech processing tasks. 
Leveraging the unique characteristic of discrete units, we reformulate (1) speech classification tasks (speech to class label), (2) sequence generation tasks (speech to label sequence), and (3) speech generation tasks (speech to speech) into a unified \emph{speech-to-unit generation} task. 
In the meantime, we propose utilizing a learnable verbalizer specifically for addressing speech classification and sequence generation tasks.
Despite its simplicity as a linear transformation, this verbalizer can effectively utilize the information encapsulated in the discrete units, bridging that rich information with the downstream labels.
The experiment results show that with the proposed method, the speech LM can solve speech classification tasks and sequence generation tasks with competitive performance compared to the ``pre-train, fine-tune'' paradigm. Also, thanks to the generative capability of the speech LMs, the proposed method can also deliver promising results on speech generation tasks, which are challenging for the fine-tuning paradigm. All the tasks are solved in a unified pipeline and with promising trainable parameter efficiency.

The advantages of the proposed unified prompt framework are as follows:
(1). We are pioneers in introducing prompt engineering to the speech domain. Our proposed method achieves results comparable to the fine-tuning approach based on self-supervised learning.
(2). Compared with the ``pre-train, fine-tune'' paradigm, our unified framework is adaptable to a wide range of speech tasks and eliminates the need for designing task-specific downstream models and loss functions. This approach not only saves considerable effort but also paves the way for a universal speech model.
(3). The learnable verbalizer boasts commendable explainability and adeptly utilizes the semantic information within the discrete units. This capacity allows for an effective linkage of that information with the labels associated with various downstream tasks.
(4). The evolution from GSLM~\cite{lakhotia2021generative} to Unit mBART~\cite{popuri22_interspeech} has significantly enhanced the performance of our prompt framework. With more advanced speech LMs coming into the stage, we anticipate these developments will elevate our methods to unprecedented levels of success.
(5). Imagine a near future where speech language models are offered in the cloud servers by major companies and widely adopted by numerous smaller businesses. In this scenario, our prompt framework utilizes discrete units, which save storage space, speed up data transmission, and potentially improve privacy. For example, discrete speech units have demonstrated a tendency to reduce speaker (timbre) information\cite{DBLP:conf/interspeech/PolyakACKLHMD21}. Therefore, employing discrete speech units can potentially mitigate privacy concerns compared to transmitting raw speech~\footnote{The extent of speaker information removal depends on the context. Recent studies~\cite{nguyen2022, kharitonov2022textless} show that when the number of discrete units increases, the retention of speaker information may become more noticeable.}.


\section{Related Works}

\subsection{Self-supervised Speech Representation and Discretization}
The exploration of speech representations through Self-Supervised Learning (SSL) objectives has evolved into a crucial research topic within the speech research area in recent years.
By utilizing different SSL pre-training tasks, the representation models can mainly be grouped into three categories: predictive models~\cite{ hsu2021hubert, chiu2022self}, contrastive models~\cite{riviere2020unsupervised, baevski2020wav2vec, jiang21_interspeech}, and generative models~\cite{chung2020generative, ling2020decoar, liu2021tera}. 
To leverage SSL representations, a common way is to build specialized downstream models on top of SSL representations and fine-tune the entire model or only the downstream models for supervised downstream tasks. 
Based on this, SUPERB~\cite{DBLP:conf/interspeech/YangCCLLLLSCLHT21} benchmarks SSL speech models with a wide variety of downstream tasks. 

Although using continuous SSL representations as features for downstream tasks can yield stronger performance~\cite{chang2023exploring}, there's a growing trend of adopting discrete speech units derived by quantizing the SSL representations~\cite{chang23b_interspeech, yang2023towards}.
A common approach involves applying the K-means algorithm to the SSL representations, quantizing them into clusters. Discrete units significantly reduce storage space and transmission bandwidth compared to raw waveforms and SSL features~\cite{chang2023exploring, chang23b_interspeech}. For instance, as discussed in ~\cite{chang23b_interspeech} and shown in Table~\ref{tab:speech_format}, a \(T\)-second 16kHz waveform in 16-bit format requires 16 \( \times \) 16,000 \( \times \) \( T \) bits for storage and transmission. In contrast, HuBERT representation with a dimension of 768 and a frame rate of 50 per second results in 6.4 times the data size using floating-point vectors (32-bit). Discrete units with 100 clusters (approximately 7 bits) and 1,000 clusters (approximately 10 bits) offer even more efficient speech data formats.

\begin{figure}[t]
    \centering
    \includegraphics[width=0.48\textwidth]{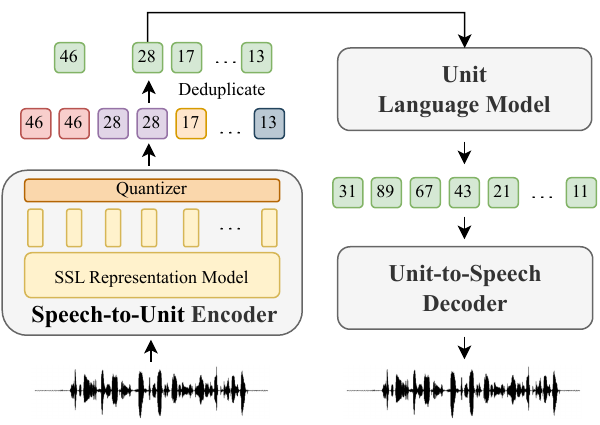}
    \caption{The textless speech LM. It consists of three components, including (1) The speech-to-unit encoder, (2) the unit language model, and (3) the unit-to-speech decoder.}
    \label{fig:textless}
\end{figure}

\begin{table}[t]
\centering
\caption{Data size for different formats of \(T\)-second speech.}
\begin{tabular}{lcc}
\toprule
Data format                & Data size (bits) & Size ratio \\ 
\toprule
Raw waveform                        & 16 \(\times\) 16000 \(\times\) \(T\)   & 1 \\ 
SSL representation                  & 32 \(\times\) 1024 \(\times\) 50 \(\times\) \(T\) & 6.4                 \\ 
HuBERT units (100 clusters)              & 7 \(\times\) 50 \(\times\) \(T\) & $~1\times10^{-3}$ 	  \\
HuBERT units (1,000 clusters)              & 10 \(\times\) 50 \(\times\) \(T\) 	& $~2\times10^{-3}$  \\
\toprule
\end{tabular}
\label{tab:speech_format}
\end{table}

\subsection{Textless Speech Language Models}
Textless speech LMs regard discrete speech units as \emph{pseudo-text} and adopt them as LM's vocabulary.
Leveraging these discrete units, speech LMs are trained to perform language modeling tasks that mirror those in the NLP field. 

As shown in Fig.~\ref{fig:textless}, in the textless speech language model, there are three components: (1) speech-to-unit encoder, (2) unit language model, and (3) unit-to-speech decoder. 
\textbf{Speech-to-unit encoder} comprises an SSL representation model, such as HuBERT~\cite{hsu2021hubert}, paired with a quantizer, like K-means. 
The continuous representation extracted by the SSL model is clustered into discrete units. 
These discrete units have shown to encapsulate rich phonetic and linguistic information, thereby effectively representing speech~\cite{lakhotia2021generative, DBLP:conf/interspeech/PolyakACKLHMD21}.
In conventional speech language models, these discrete units undergo a \emph{deduplication} process, which removes consecutive repeated units to form a more compact sequence of tokens for language modeling. 
The \textbf{unit language model} is an LM that performs generative language modeling based on the discrete units.
For instance, in GSLM~\cite{lakhotia2021generative}, the unit language model conducts the next-token-prediction task akin to GPTs~\cite{radford2018improving, radford2019language}. Unit mBART performs the denoising sequence reconstruction task similar to the BART model~\cite{DBLP:conf/acl/LewisLGGMLSZ20}.
The \textbf{unit-to-speech decoder} is responsible for transforming the generated discrete unit sequences back into continuous speech signals. The architecture is akin to the conventional speech synthesis models~\cite{shen2018natural, kong2020hifi}  that train on the unit sequence and speech signal. 

In addition to GSLM and Unit mBART, there are other notable speech language models such as AudioLM~\cite{borsos2023audiolm}, TWIST~\cite{DBLP:conf/nips/HassidRNGCKCDSD23}, and SPECTRON~\cite{nachmani2023lms}. These models bring additional complexity and advancements to the field; however, they are not currently fully open-sourced. As the development of speech LMs continues to evolve, there is significant potential for our framework to be expanded and utilized more extensively in future research and applications.

\begin{figure}[t]
    \centering
    \includegraphics[width=0.48\textwidth]{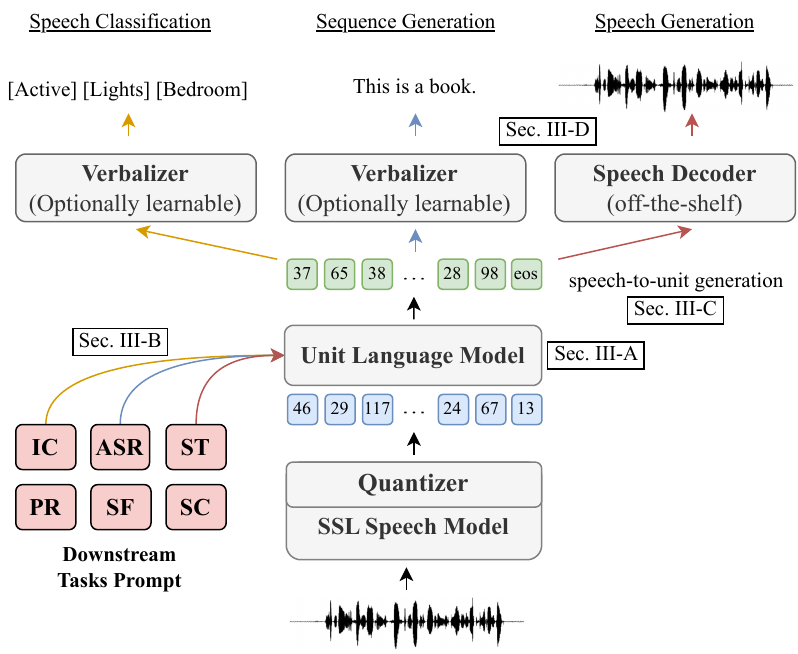}
    \caption{An overview of the proposed framework, where all downstream tasks are treated as speech-to-unit generation processes.
The generation of units is directed by the task-specific prompts that guide the unit language model. A verbalizer or speech
decoder then bridges the gap between the generated units and the corresponding downstream labels.}
    \label{fig:unit-generation}
\end{figure}

\subsection{Prompting and Reprogramming in Speech Processing} 
\label{sec:prompt}

This journal paper is an extension of our previous work~\cite{chang22e_interspeech}, where we explored the concept of prompting on speech LM, particularly GSLM. Previous work~\cite{chang22e_interspeech} showed promising results in speech classification tasks such as spoken command recognition and intent classification and demonstrated better parameter efficiency compared to the ``pre-train, fine-tune'' paradigm. However, despite achieving notable results in sequence generation tasks like ASR and slot filling, its performance still lags behind the fine-tuning method.
In this paper, we further explore an advanced encoder-decoder speech LM, Unit mBART, across a broader range of speech processing tasks. This includes a more diverse set of speech classification tasks, as well as speech generation tasks. The results are more promising: (1) Prompting Unit mBART achieves competitive performance in sequence generation tasks and (2) Prompting Unit mBART is well-suited for speech generation tasks, thereby establishing a unified prompting framework for various speech processing tasks. 
Additionally, compared to our previous work, we introduce a learnable verbalizer in this paper to bridge the gap between discrete units and downstream task labels, enhancing both explainability and performance.

WavPrompt~\cite{gao22e_interspeech} is also a pioneer in studying the prompting paradigm in speech processing. WavPrompt consists of a text LM, GPT-2~\cite{radford2019language}, and an audio encoder, wav2vec 2.0~\cite{baevski2020wav2vec}. The text LM is prompted with audio embeddings and text questions to perform few-shot speech understanding tasks. In contrast to SpeechPrompt, which uses textless speech LM for various speech processing tasks, WavPrompt employs a text LM and performs limited speech understanding tasks. 

On the other hand, the work~\cite{peng23d_interspeech} studies hand-crafted prompts for a speech recognition model, Whisper~\cite{radford2023robust}, for various speech recognition tasks. The backbone model, Whisper, is trained using large-scale speech-text paired data. In contrast, our work prompts a textless speech LM, and we not only focus on speech recognition, a type of sequence generation task, but also explore speech generation tasks.


Another branch of utilizing a pre-trained model's capability for different tasks is \emph{model reprogramming}~\cite{elsayed2018adversarial, chen2024model}. 
In~\cite{yang2021voice2series, yen23_interspeech}, the input data (target domain) are first transformed with a task-specific function to become the reprogrammed data. The pre-trained acoustic model is then capable of generating labels for this reprogrammed data.
These labels (source domain) are then mapped to the classes of downstream tasks (target domain) by a mapping function. 
This mapping function serves the same role as the verbalizer in the prompting method and is usually a random mapping in the reprogramming literature.
We also adopt the idea of reprogramming a foundation model for solving various tasks. 
For example, in speech classification tasks and sequence generation tasks, the speech LM is prompted/reprogrammed to adapt to the distribution of the target domain (the class label and the transcription).

\begin{table}[t]
\centering
\caption{Notation Table}
\label{tab:notation}
\begin{tabular}{c|p{0.79\linewidth}}
\toprule
\textbf{Symbol} & \textbf{Description} \\ 
\toprule
\( u \) & Unit in the unit sequence \\
\( \bm{u}^x \) & Discretized speech, source unit sequence \\
\( \bm{u}^y \) & Generated target unit sequence \\
\( C \) & Context, including the input discretized speech, sequence of units before the current unit and the task prompts \\
\( \bm{z}_{tj} \) & Logit for \( j \)-th unit at timestep \( t \) \\
\( P(u_j | C_t) \) & Probability of unit \( u_j \) at timestep \( t \) given context \( C_t \) \\
\( \mathcal{E} \) & Encoder in encoder-decoder unit LM \\
\( \mathcal{D} \) & Decoder in decoder-only or encoder-decoder unit LM \\
\( \bm{e}(u) \) & Unit LM's vocabulary embedding vector for a unit \( u \) \\
\( V \) & Vocabulary set \( \{u_1, u_2, \ldots, u_{|V|}\} \) \\
\( \bm{g}^{(i)} \) & Hidden representation input to the \( i \)-th layer of encoder \\
\( \bm{h}^{(i)} \) & Hidden representation input to the \( i \)-th layer of decoder \\
\( T \) & Sequence length of encoder's hidden representation \\
\( T' \) & Sequence length of decoder's hidden representation \\
\( \bm{p} \) & Trainable prompt sequence \([p_1, \ldots p_l] \) with prompt length \(l \)\\
\( \bm{y}\) & Downstream label sequence \([y_1, \ldots y_{T'}]\) \\
\( |Y|\) & Number of classes in the downstream task \\

\toprule
\end{tabular}
\vspace{-0.3cm}
\end{table}

\section{Method}
\label{sec:method}

The overview of the proposed framework is depicted in Fig.~\ref{fig:unit-generation}. The input speech waveform is encoded into a sequence of discrete units using an SSL speech model and a quantizer. The unit LM (Section~\ref{sec: UnitLM}) then takes this unit sequence and performs conditional generation based on the task-specific prompts. The design of task-specific prompts will be illustrated in Section~\ref{sec: prompt}. The prompts steer the unit LM to solve the downstream speech processing task, which is reformulated into a speech-to-unit generation task as discussed in Section~\ref{sec: Speech-to-Unit Generation}. The resulting unit sequence is transformed into the downstream task's target through a verbalizer (for speech classification and sequence generation tasks) or through a pre-trained speech decoder (for speech generation tasks) as discussed in Section~\ref{sec:verbalizer}. 
The notations used in the section are listed in Table~\ref{tab:notation}.

\subsection{Unit Language Models}
\label{sec: UnitLM}
\begin{figure*}[t]
    \centering
    \includegraphics[width=0.98\textwidth]{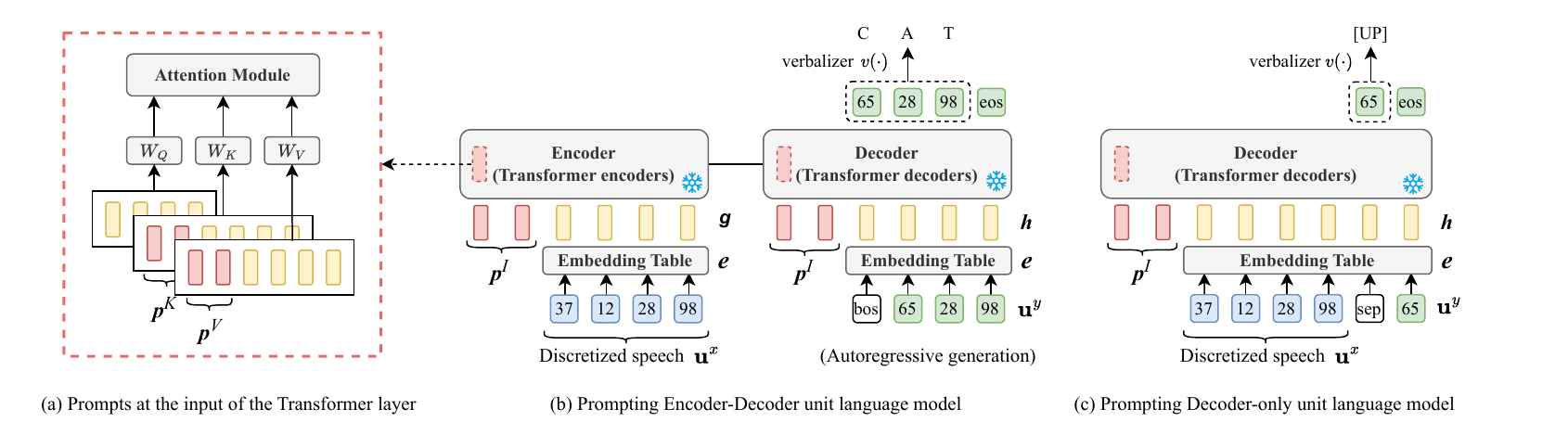}
    \caption{An overview of the proposed framework, where all downstream tasks are treated as speech-to-unit generation processes. The generation of units is directed by the task-specific prompts that guide the unit language model. A verbalizer or speech decoder then bridges the gap between the generated units and the corresponding downstream labels.}
    \label{fig:prompt-tuning}
\end{figure*}

This subsection explains the backbone unit language model in our prompt framework.
As shown in Fig.~\ref{fig:prompt-tuning}, these unit LMs receive discretized speech units sequence $\bm{u}^x$ and trainable prompts $\bm{p}$ as inputs, subsequently using them to generate target unit sequence $\bm{u}^y$ for downstream speech processing tasks.


Without loss of generality, in this paper, we investigate two variants of widely-adopted unit LMs based on Transformers~\cite{vaswani2017attention}: (1) The decoder-only unit LM that mimics the GPT architecture~\cite{radford2018improving}, and (2) The encoder-decoder unit LM that mirrors the BART language model~\cite{DBLP:conf/acl/LewisLGGMLSZ20}.
Both model types employ a causal decoder and are characterized as autoregressive LMs, enabling the capability to generate outputs of varying lengths. Specifically, the probability of each unit $u^y_t \in \bm{u}^y$ generated by the model at the timestep $t$ is conditioned on the preceding context, denoted by $C_t$. 
The context $C_t$ includes the input discretized source speech $\bm{u}^x$, the task prompts $\bm{p}$, and the units $\bm{u}^y_{<t}$ generated preceding the timestep $t$ in the autoregressive process. Formally, the autoregressive model generates the probability of a unit \( u_j \) within a vocabulary \( V = \{u_1, u_2, \ldots, u_{|V|}\} \) at timestep \( t \) given the context \( C_t \) as:

\begin{equation}
    P(u_j | C_t) = \frac{e^{\bm{z}_{tj}}}{\sum_{k=1}^{|V|} e^{\bm{z}_{tk}}},    
\end{equation}

where \( \bm{z}_{tj} \in \mathbb{R}^{|V| \times 1}\) is the logit for the \( j \)-th unit at timestep \( t \), and the denominator is the sum of exponentiated logits for all units at that timestep. 

\subsubsection{Encoder-Decoder Unit LM}
The encoder-decoder unit LM includes the encoder \( \mathcal{E} \) and decoder \( \mathcal{D} \) based on Transformer. The discretized speech is first processed by the encoder \( \mathcal{E} \) to form part of the enriched context that the decoder \( \mathcal{D} \) performs cross-attention on to guide the generation of the discrete units.
The encoder \( \mathcal{E} \) is composed of multiple layers that process the input unit sequence:
\begin{equation}
\label{eq:input_embed_}
    \bm{g}^{(1)} = [\bm{e}(u_1^x), \bm{e}(u_2^x), \ldots, \bm{e}(u_{T}^x)],    
\end{equation}
where \( T \) is the sequence length and $ \bm{e}(\cdot): \mathbb{Z} \mapsto \mathbb{R}^d$ denotes the vocabulary embedding table, which transforms a discrete unit $u \in \mathbb{Z}$ into its corresponding embedding vector $\bm{e}(u) \in \mathbb{R}^d$, and $d$ is the embedding dimension.
In the encoder, the \( i \)-th layer receives hidden representation \( \bm{g}^{(i)} = [g_1^{(i)}, g_2^{(i)}, \ldots, g_T^{(i)}] \) as input and outputs \( \bm{g}^{(i + 1)} \). 
The decoder layers operate similarly, with each taking input $\bm{h}^{(i)} = [h_1^{(i)}, h_2^{(i)}, \ldots, h_{T'}^{(i)}]$, and outputs $\bm{h}^{(i+1)}$, where \( T' \) represents the decoder sequence length, which increases incrementally during the autoregressive process.

\subsubsection{Decoder-only Unit LM}
In the decoder-only LM, the model lacks the encoder and relies solely on the decoder \( \mathcal{D} \), which functions in an analogous fashion to the encoder-decoder setup but without the encoder's guidance.  Without the encoder, the discretized source speech $\bm{u}^x$ is integrated at the beginning of the sequence, serving as the initial context for the decoder to predict the subsequent units. A separation token $\langle sep \rangle$ is inserted in between the source unit sequence $\bm{u}^x$ and the generated units $\bm{u}^y$. Therefore, for each timestep $t$ in the autoregressive process, the input to the decoder \( \mathcal{D}\) is:

\begin{equation}
\label{eq:input_embed}
    \bm{h}^{(1)} = [\bm{e}(\bm{u}^x), \bm{e}(\langle sep \rangle), \bm{e}(u^y_1), \ldots, \bm{e}(u^y_{<t})].
\end{equation}

\subsection{Prompt Tuning}
\label{sec: prompt}
As depicted in Fig.~\ref{fig:unit-generation}, the speech LM is capable of performing predefined speech tasks when provided with various types of prompts. In this subsection, we will elaborate on the process of prompt design.

Prompting employs task-specific templates, known as prompts, to steer the generation process of the LM.
This technique involves freezing the LM's parameters while integrating prompts as part of the input.
Our method, inspired by the prompt tuning approaches~\cite{li2021prefix, lester2021power}, is implemented in two positions: (1) at the input of the unit LM, termed \emph{input prompt tuning}, and (2) at the input of each Transformer layer, termed \emph{deep prompt tuning}.

\subsubsection{Input Prompt Tuning}
Inspired by the method in \cite{lester2021power}, input prompt tuning prepends continuous prompt vectors at the LM's input. Specifically, the prompts are prepended at the embedding sequence of the first layer's input $\bm{h}^{(1)}$ (and $\bm{g}^{(1)}$ for Encodoer-Decoder model):
\begin{align}
    \bm{h}^{(1)} &\leftarrow Concat(\bm{p}^{I}, \bm{h}^{(1)}), \\
    \bm{g}^{(1)} &\leftarrow Concat(\bm{p}^{I}, \bm{g}^{(1)}),
\end{align}
where \( \bm{p}^{I} = [p^{I}_1, p^{I}_2, \ldots, p^{I}_l] \) represents a series of prompt vectors $p \in \mathbb{R}^d$ at the input of the unit LM, with \( l \) indicating the prompt length.

\subsubsection{Deep Prompt Tuning}
Inspired by prefix-tuning~\cite{li2021prefix}, deep prompt tuning involves concatenating prompt vectors at the input of the Transformer layer. 
Specifically, it modifies the input of the attention modules to guide the forward process of the LM. 
The self-attention module at the beginning of each transformer layer takes the Query ($Q$), Key ($K$), and Value ($V$) as input:
\begin{equation}
    Attn(Q, K, V) = softmax\left(\frac{QK^T}{\sqrt{d_k}}\right)V,
\end{equation}
where \( \sqrt{d_k} \), the square root of the dimensionality of the key vectors, scales the dot product to ensure normalization of the attention weights by the softmax function.
For self-attention, the matrices $Q$, $K$, and $V$ are projections of the same input \( \bm{g} \) or \( \bm{h} \) transformed by the weight matrices \( W_Q \), \( W_K \), and \( W_V \), respectively.
Trainable prompt vectors are prepended to the input of each transformer layer, affecting both Key ($K$) and Value ($V$) matrices in the attention mechanism:
\begin{align}
    K &\leftarrow Concat(\bm{p}^K, \bm{h})W_K, \\
    V &\leftarrow Concat(\bm{p}^V, \bm{h})W_V,
\end{align}
where $\bm{p}^K = [p^{K}_1, p^{K}_2, \ldots, p^{K}_l]$ and $\bm{p}^V = [p^{V}_1, p^{V}_2, \ldots, p^{V}_l]$ are series of trainable prompt vectors for key and value, respectively, and has the same prompt length $l$ as $\bm{p}^{I}$.

Similar adjustments are applied to the encoder's representation $\bm{g}$ for encoder-decoder unit LM. It is crucial to note that throughout the prompt tuning process, only the prompt vectors are trainable. The embedding table and the unit LM remain fixed.

\subsection{Speech-to-Unit Generation}
\label{sec: Speech-to-Unit Generation}
In this paper, we focus on leveraging the generative capabilities of autoregressive speech LMs to handle various downstream tasks. 
Specifically, we recast speech processing tasks, including speech classification, sequence generation, and speech generation, into a unified \emph{speech-to-unit generation task}.
In this approach, speech LM takes discretized speech as input and generates a sequence of discrete units corresponding to the intended output for the task at hand. 

In \textit{sequence generation} tasks, like automatic speech recognition (ASR), the model generates a unit sequence $\bm{u}^y = [u^y_1, ..., u^y_{T'}, \langle eos \rangle]$. Each unit $u_t^y$ represents a discrete token corresponding to the character $y_t$ in the target character sequence $\bm{y} = [y_1, ..., y_{T'}]$. The mapping from units to characters is facilitated by the verbalizer, detailed in Section~\ref{sec:verbalizer}.
For \textit{speech classification} tasks like spoken command recognition (SCR), which involve single-label classification, the model's goal is to classify an utterance into a predefined category. Instead of directly predicting a label $y_1$, it generates a unit sequence $\bm{u}^y = [u^y_1, \langle eos \rangle]$, where $u_1^y$ will be transformed into the label $y_1$. 
In \textit{speech generation} tasks, the generated unit sequence can be synthesized back into the target speech signal using an off-the-shelf unit-to-speech decoder. 
Notably, the autoregressive nature of speech LMs allows them to handle varying label lengths across different tasks, thus enabling a unified framework.

\subsection{Verbalizer and Speech Decoder}
\label{sec:verbalizer}
Within the prompting paradigm, the \emph{verbalizer}~\cite{DBLP:conf/naacl/SchickS21, DBLP:conf/eacl/SchickS21} \( v(\cdot) \) is a label-mapping module, which establishes the connection between the downstream task labels and the LM's vocabulary. For speech LM, the vocabulary is the discrete units. The verbalizer can adopt various forms, including random mapping~\cite{schick2020automatically, min2022rethinking, yen23_interspeech} and heuristic methods~\cite{chang22e_interspeech}, we refer to this as ``fixed verbalizer'' since the mapping is pre-defined and does not include updates. On the other hand, to generate speech signal, a speech decoder is employed to synthesize waveform from discrete unit sequence~\footnote{While the term ``verbalizer'' is usually associated with the concept of ``speaking'', we use it here to refer to the label-mapping module in line with the prompting paradigm in NLP~\cite{DBLP:conf/eacl/SchickS21, hu2022knowledgeable, cui2022prototypical}. The verbalizer connects the LM’s output (discrete tokens or probability distributions) to the labels of downstream tasks, facilitating task-oriented responses. On the other hand, the ``speech decoder'' is responsible for transforming the LM's outputs into audible speech.
}.

\subsubsection{Fixed Verbalizer}
The fixed verbalizer establishes a static mapping between the downstream task label and a unique unit. 
For example, in the ASR task, it might map the character ``a'' to ``unit 28'' and ``b'' to ``unit 72.'' In spoken command recognition (SCR), it could map the command ``[UP]'' to ``unit 65,'' following either a random mapping or a frequency-based approach~\cite{chang22e_interspeech}.
Once established, this mapping remains static without further learning or adaptation. 
In practice, with a fixed verbalizer, the most probable unit at each timestep \( t \) is selected and directly converted to the downstream task's label \( y_t \).



\subsubsection{Speech Decoder}
For speech generation tasks, where the target output is a speech signal rather than a sequence of labels, the discrete units can be synthesized back into speech signals using a pre-trained, off-the-shelf unit-to-speech decoder. 
This speech decoder is self-supervised and trained with pairs of discrete units and their corresponding speech. In this work, we employ a speech decoder that corresponds to the given unit LM, as illustrated in Fig.~\ref{fig:textless}.

\begin{figure}[t]
    \centering
    \includegraphics[width=0.49\textwidth]{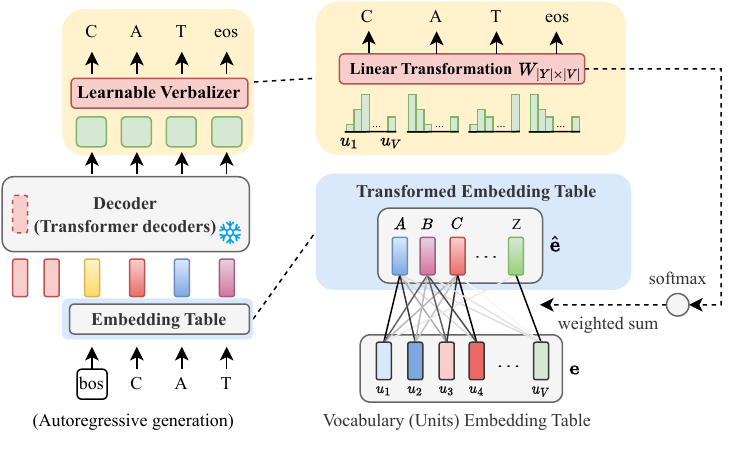}
    \caption{Illustration of the learnable verbalizer. The logits are transformed into labels for the downstream task through a linear transformation. Furthermore, the original vocabulary embeddings are converted into class-specific embeddings using weighted transformations, aligning them more closely with the downstream task.
    }
    \label{fig:verbalizer}
\end{figure}

\subsection{Learnable Verbalizer}
Fixed verbalizers can lead to subpar performance in speech processing tasks because, unlike the distinct semantic meaning present in NLP vocabulary, the vocabulary of discrete speech units lacks clear semantic meanings.
 To address this, we introduce a learnable verbalizer coupled with a novel input transformation (Fig.~\ref{fig:verbalizer}) that aligns the discrete units with the downstream task labels more meaningfully.
In a learnable verbalizer, the mappings are determined by a learnable linear transformation matrix $W \in \mathbb{R}^{|Y| \times |V|}$, where \( |V| \) is the size of the original LM's vocabulary, and \( |Y| \) is the number of classes in the downstream task. This matrix is applied to the logits vector \( \bm{z}_t \in \mathbb{R}^{|V| \times 1}\) to produce a transformed logits vector \( \hat{\bm{z}}_t \in \mathbb{R}^{|Y| \times 1}\) over the downstream task labels:

\begin{equation}
\label{eq:learn_verb}
\bm{\hat{z}}_t = W \cdot \bm{z}_t
\end{equation}

Following this transformation, the label \( y_t \) is sampled from the transformed logits:
\begin{equation}
y_t = \underset{y}{\mathrm{argmax}} \ P(y | \bm{\hat{z}}_t),
\end{equation}
where \( P(y | \bm{\hat{z}}_t) \) is the probability of class \( y \) given the transformed logits \( \bm{\hat{z}}_t \) at timestep \( t \).

To facilitate autoregressive processing and incorporate the predicted downstream tasks' labels as input to the unit LM, we propose an input transformation mechanism, which is coupled with the learnable verbalizer matrix \( W \).  This mechanism transforms the original vocabulary embeddings into new embeddings suitable for the downstream task's labels.

Mathematically, the transformed embedding for a given class \( y \), denoted as \( \bm{\hat{e}}(y) \), is computed as a weighted sum of the original vocabulary embeddings \( \bm{e}\). 
Let \( W_{y:} \) denote the \( y \)-th row of the matrix \( W \), where the \( i \)-th element in \( W_{y:} \) represents the learned weight of the \( i \)-th unit contributing to the class \( y \). \( W_{y:} \) is then input into the softmax function with the temperature parameter \( \tau \), transforming the weights into probabilities. The formula is expressed as:

\begin{equation}
\bm{\hat{e}}(y) = \sum_{i=1}^{|V|} \left( softmax\left( \frac{W_{y:}}{\tau} \right) \right)_i \cdot \bm{e}(u_i)
\end{equation}

In this formulation, \( \bm{\hat{e}}(y) \) signifies the newly generated embedding for class \( y \), tailored to the demands of the downstream task. The process effectively creates class-specific embeddings by aggregating the original embeddings, each weighted according to the transformed softmax outputs. This method not only preserves the intrinsic properties of the original vocabulary embeddings but also aligns them more closely with the target classes of the downstream task, thereby enhancing the model's adaptability and effectiveness in handling varied speech processing applications.

The learnable verbalizer offers improved explainability, effectively utilizing the information in the discrete units, which will be discussed in Sec.~\ref{sec:verb_ana}. Meanwhile, it preserves parameter efficiency in the prompting paradigm. For instance, in the ASR task featuring 28 classes and a Unit mBART model with 1,000 units, the verbalizer necessitates fewer than 30,000 learnable parameters. 
\begin{table}[t]
\centering
\caption{The downstream tasks performed in this paper, including speech classification, sequence generation, and speech generation tasks. Language abbreviations are in ISO 639-1 format. $\bm{N}_{\text{class}}$: Number of classes for each downstream task. $\overline{|\bm{u}^x|}$: average discrete unit length of the utterance. $\overline{|\bm{u_{de}}|}$: average deduplicated discrete unit length of the utterance. $\overline{|\bm{y}|}$: average label length.}
\renewcommand{\arraystretch}{0.9} 
\setlength{\tabcolsep}{5pt} 
\begin{tabular}{ll|c|c|c|c|c}
\toprule
\textbf{Task} & \textbf{Dataset} & \textbf{Language} & $\bm{N}_{\text{class}}$ & $\overline{|\bm{u}^x|}$ & $\overline{|\bm{u_{de}}|}$ & $\overline{|\bm{y}|}$ \\
\midrule \midrule 
\multicolumn{7}{c}{\textit{Speech Classification}} \\
\midrule
\multirow{5}{*}{SCR} & Google SC v1    & en & 12 & 48 & 25 & 1 \\
                     & Arabic SC       & ar & 16 & 41 & 25 & 1 \\
                     & Lithuanian SC   & lt & 15 & 51 & 28 & 1 \\
                     & DM-SC           & zh & 19 & 169 & 68 & 1 \\
                     & Grabo SC        & nl & 36 & 132 & 71 & 1 \\
\midrule
IC                   & Fluent SC       & en & 24 & 115 & 61 & 3 \\
\midrule
SD & Mustard++ & en & 2  & 229 & 128 & 1 \\
\midrule
AcC                  & AccentDB        & en & 9  & 205 & 91 & 1 \\
\midrule
LID                  & Voxforge        & \begin{tabular}[c]{@{}c@{}}en, es,\\ fr, de,\\ ru,\, it\end{tabular} & 6 & 392 & 231 & 1 \\
\midrule
VAD                  & \begin{tabular}[c]{@{}c@{}}Google SC v2 /\\ FreeSound \end{tabular}       & \begin{tabular}[c]{@{}c@{}}en /\\ audio \end{tabular} & 2 & 31 & 16 & 1 \\

\midrule \midrule
\multicolumn{7}{c}{\textit{Sequence Generation}} \\
\midrule
ASR                  & LibriSpeech     & en & 28 & 591 & 355 & 172 \\
\midrule
PR                   & LibriSpeech     & en & 71 & 591 & 355 & 116 \\
\midrule
SF                   & AudioSNIPS      & en & 107 & 142 & 96 & 53 \\
\midrule\midrule 
\multicolumn{7}{c}{\textit{Speech Generation}} \\
\midrule
ST                   & CoVoST2         & en, es & $|V|$ & 305 & 167 & 120 \\
\midrule
SC                   & LJSpeech        & en, es & $|V|$ & 328 & 199 & 199 \\
\bottomrule
\end{tabular}
\label{tab:dataset}
\end{table}

\section{Experimental Setup}
In this work, we compare the ``pre-train, fine-tune'' paradigm with the prompting paradigm for speech processing across three types of tasks: (1) speech classification tasks, (2) sequence generation tasks, and (3) speech generation tasks.
The used dataset and the basic statistics are presented in Table ~\ref{tab:dataset}.
\subsection{Tasks and Datasets}
\subsubsection{Speech Classification Tasks}

\textbf{\\Speech Command Recognition (SCR):} The task is to recognize which keyword is presented in a given utterance. 
We adopted the Google Speech Commands dataset~\cite{warden2018speech} and low-resource datasets in different languages. 
These include Grabo Speech Commands (Grabo-SC)~\cite{DBLP:conf/interspeech/TianG20}, Lithuanian Speech Commands (LT-SC)~\cite{kolesau2020unsupervised}, Dysarthric Mandarin Speech Commands (DM-SC)~\cite{lin2021speech}, and Arabic Speech Commands (AR-SC)~\cite{benamer2020database}.

\noindent\textbf{Intent Classification (IC):} This task classifies utterances into predefined classes to determine the intent of speakers. We used the Fluent Speech Commands dataset~\cite{fluent}, where each utterance has three labels: action, object, and location.

\noindent\textbf{Sarcasm Detection (SD):} This task aims to determine if an utterance is sarcastic. We employed the Mustard++ dataset~\cite{DBLP:conf/lrec/RayMNB22}.

\noindent\textbf{Accent Classification (AcC):} This task involves classifying accents within the same language. We utilized the AccentDB Dataset~\cite{ahamad2020accentdb}, which includes four Indian-English accents, four native English accents, and one metropolitan Indian-English accent.

\noindent\textbf{Language Identification (LID):} The objective of this task is to recognize the language present in a given utterance. 
We utilized the Voxforge Dataset~\cite{maclean2018voxforge}, which comprises utterances in six different languages.

\noindent\textbf{Voice Activity Detection (VAD):} This task is to determine whether a segment of an utterance contains human speech or is just background noise or silence. Following MarbleNet~\cite{jia2021marblenet}, we used Google Speech Commands v2~\cite{warden2018speech} as speech data and FreeSound dataset~\cite{DBLP:conf/ismir/FonsecaPFFBFOPS17} as background noise data. We refer to this mixed dataset as GFSound.

The evaluation metric for speech classification tasks mentioned above is accuracy.

\subsubsection{Sequence Generation Tasks}
\noindent\textbf{\\Automatic Speech Recognition (ASR):} The task is to transcribe an utterance into text (character sequence). We utilized the LibriSpeech~\cite{DBLP:conf/icassp/PanayotovCPK15} train-clean-100 dataset for training and the test-clean dataset for testing. The evaluation metrics are word error rate (WER) and character error rate (CER).

\noindent\textbf{Phoneme Recognition (PR):} The task involves transcribing an utterance into a phoneme sequence. We utilized LibriSpeech  train-clean-100 and test-clean datasets for training and testing.
The evaluation metric is phoneme error rate (PER).

\noindent\textbf{Slot Filling (SF):} In the slot filling task, models are expected not only to recognize the spoken content but also to decode the associated slot type. Specifically, the slot type is decoded in conjunction with the transcription in a sequence generation approach. We adopted AudioSNIPS dataset~\cite{DBLP:conf/icassp/LaiCL0G21} and the evaluation metrics are character error rate (CER) and F1 score.\\

\vspace{-0.1cm}
\subsubsection{Speech Generation Tasks}
\noindent\textbf{\\Speech Translation (ST):} ST is the process of converting speech signals from the source language into speech in the target language, enabling communication between individuals who speak different languages. 
We utilize the CoVoST2~\cite{wang21s_interspeech} Es-En dataset. This dataset comprises parallel text data for Spanish (Es) and English (En) translations. Following ~\cite{popuri22_interspeech}, we utilize a single-speaker TTS system~\footnote{https://huggingface.co/espnet/kan-bayashi\_ljspeech\_vits} to synthesize the speech of the target language. We utilize an off-the-shelf ASR system~\footnote{``Wav2vec2\_large\_lv60k'' with CTC decoder available on PyTorch} to transcribe the generated speech and calculate the BLEU with sacrebleu~\footnote{https://github.com/mjpost/sacrebleu}.
We perform Mean Opinion Score (MOS) prediction with TorchAudio-Squim~\cite{kumar2023} to assess the naturalness of the generated speech~\footnote{TorchAudio-Squim utilizes non-matching reference (NMR) for MOS prediction. For assessing each utterance, we randomly sample one clean speech from the original dataset as the reference.}.


\noindent\textbf{Speech Continuation (SC):} SC aims to generate coherent continuation of a given speech input while preserving the semantic context.
In the experiment, we adopt LJSpeech \cite{ljspeech17}, which contains approximately 24 hours of English speech from a single speaker. 
We divided the LJSpeech dataset into training, validation, and testing subsets. 
Within these subsets, we designated each utterance's initial $r$ fraction as the seed segment for the speech continuation tasks. 
We refer to the value of $r$ as the \emph{conditional ratio}. Given this seed segment, our model aims to generate a coherent continuation of the speech. Following ST, the generated speech is first transcribed into text, after which Perplexity (PPX)~\footnote{Evaluated with a pre-trained LM, ``transformer\_lm.wmt19.en'', available on Fairseq.} and Auto-BLEU~\cite{lakhotia2021generative} are evaluated. We report MOS prediction results to assess the naturalness of the speech and evaluate the speaker similarity (SIM) between the seed segment and the generated speech. SIM is obtained by computing the cosine similarity between speaker embeddings, derived from the Resemblyzer package~\footnote{https://github.com/resemble-ai/Resemblyzer}.


\begin{table*}[t]
    \centering
    \caption{Performance comparison on speech classification tasks for the ``pre-train, fine-tune paradigm'' (\textbf{FT}) and the ``prompting paradigm'' (\textbf{PT}). The evaluation metric is accuracy.
    \textnormal{\textbf{HuBERT~+~Expert}} and \textnormal{\textbf{mHuBERT~+~Expert}}: Building an expert downstream model on top of the SSL speech model and fine-tuning the expert model. \textnormal{\textbf{GSLM$_{\text{fixed}}$}} and \textnormal{\textbf{Unit mBART$_{\text{fixed}}$}}: Prompting the speech language model with a fixed verbalizer. \textnormal{\textbf{GSLM$_{\text{learn}}$}} and \textnormal{\textbf{Unit mBART$_{\text{learn}}$}}: Prompting the speech LM with a learnable verbalizer. In the PT scenarios, around 0.15M trainable parameters are included; while FT adopts around 0.2M parameters.}
    \resizebox{\linewidth}{!}{
    \begin{tabular}{cc|ccccc|c|c|c|c|c}
        \toprule
        \multicolumn{12}{c}{Speech Classification Tasks (full dataset setting)} \\
        \midrule
        \midrule
        \multirow{2}{*}{Paradigm} & \multirow{2}{*}{Scenario} & \multicolumn{5}{c|}{SCR} & IC & SD & AcC & LID & VAD \\
        & & Google-SC & AR-SC & LT-SC & DM-SC & Grabo-SC  & Fluent SC & Mustard++ & AccentDB & Voxforge & GFSound \\
        \midrule
        FT & HuBERT~+~Expert & \textbf{94.88} & 98.38 & 92.86 & 52.14 & 91.44 & 93.18 & 61.72 & \textbf{99.53} & \textbf{97.73} & \textbf{98.55} \\
        \multirow{2}{*}{PT} & GSLM$_{\text{fixed}}$ & 94.50 & \textbf{99.70} & \textbf{93.20} & 74.30 & 92.40 & \textbf{98.76} & 63.33 & 78.90 & 90.90 & 96.60 \\
         & GSLM$_{\text{learn}}$ & 94.71 & 99.19 & 92.86 & \textbf{74.36} & \textbf{95.76} & 98.58 & \textbf{65.83} & 80.02 & 87.69 & 97.01 \\
        \midrule
        \midrule
        FT & mHuBERT~+~Expert & 93.59 & 99.46 & \textbf{91.84} & 64.96 & 89.92 & 93.57 & 60.42 & \textbf{93.78} & 98.23 & \textbf{98.40} \\
        \multirow{2}{*}{PT} & Unit mBART$_{\text{fixed}}$ & 93.99 & 96.22 & \textbf{91.84} & 64.96 & \textbf{97.11} & \textbf{97.81} & \textbf{63.33} & 88.71 & \textbf{98.81} & 97.26 \\
         & Unit mBART$_{\text{learn}}$ & \textbf{94.45} & \textbf{99.73} & \textbf{91.84} & \textbf{77.78} & 95.07 & \textbf{97.81} & \textbf{63.33} & 87.38 & 98.13 & 97.43 \\
        \bottomrule
    \end{tabular}
    }
\label{tab:sc_full}
\end{table*}

\vspace{-0.2cm}
\subsection{Model and Training Setup}
We compare the ``pre-train, fine-tune'' paradigm with the prompting paradigm to assess whether prompting can achieve competitive performance while also providing parameter efficiency and other associated benefits as discussed in Table.~\ref{tab:paradigm_comparison}.

\subsubsection{Prompting Paradigm}

We explore two types of speech LMs within the prompting paradigm: the decoder-only \emph{Generative Spoken Language Model (GSLM)}~\cite{lakhotia2021generative} and the encoder-decoder model \emph{Unit mBART}~\cite{popuri22_interspeech}. 
GSLM is pre-trained using a next token prediction task on discrete units obtained by quantizing the 6-th layer of HuBERT representations into 100 clusters. The GSLM paper~\cite{lakhotia2021generative} considered different settings, including various SSL models and cluster numbers. This setting is selected for its superior performance. 
GSLM consists of 12 Transformer-decoder layers, each with 16 attention heads, an embedding size of 1024, and a feedforward network (FFN) size of 4096, totaling 150 million parameters. It is pre-trained on HuBERT discrete units derived from a clean 6k-hour sub-sample of the Libri-Light dataset~\cite{kahn2020libri}. 
On the other hand, Unit mBART is pre-trained on a multilingual denoising task using discrete units derived from quantizing the 11-th layer of mHuBERT representations into 1,000 clusters. Unit mBART includes 12 Transformer-encoder layers and 12 Transformer-decoder layers, each with an embedding size of 1024, an FFN dimension of 4096, and 16 attention heads, totaling 353 million parameters. The embedding tables of the encoder and decoder share the same parameters. Unit mBART is pre-trained on mHuBERT discrete units obtained from VoxPopuli with 16k hours for Spanish, 14k hours for English, and Libri-Light with 60k hours for English.


In our experiments, we set the prompt length \( l = 5 \) for GSLM on speech classification tasks and \( l = 3 \) for Unit mBART.
In sequence generation tasks, the prompt lengths are \( l = 180 \) for GSLM and \( l = 50 \) for Unit mBART.
For speech generation tasks, we adopt a prompt length of \( l = 200 \) for Unit mBART and a prompt length of \( l=180 \) for GSLM. 
The prompt length is a hyperparameter that can be adjusted for different numbers of trainable parameters. 
Since the architecture of both models is different, the positions in which the prompts can be inserted differ; notably, Unit mBART has an extra encoder for this purpose. 
Consequently, we have employed different prompt lengths for the two models with the aim of achieving competitive performance while maintaining a comparable number of trainable parameters to compare with the ``pre-train, fine-tune'' paradigm. Specifically, for GSLM in sequence and speech generation tasks, we used a prompt length of 180, which was determined to provide effective results based on the previous study~\cite{chang22e_interspeech}. Conversely, for Unit mBART, we aimed to showcase its parameter efficiency in speech classification and sequence generation tasks, as well as demonstrate feasibility in speech generation tasks. Therefore, we adopted specific prompt lengths accordingly.

We used random mapping for the fixed verbalizer. We adopt random mapping instead of a heuristic frequency-based approach~\cite{chang22e_interspeech} for two reasons: (1) In the preliminary study, we do not observe significant performance improvement when utilizing the heuristic method. (2) In the few-shot learning scenario, the statistics of the discrete units are inadequate. For the learnable verbalizer, we set the softmax's temperature \( \tau = 0.01\) in the input transformation.

For prompt tuning in each task, the prompt vectors are randomly initialized. We use the Adam optimizer with $\beta$ parameters set to (0.9, 0.98) and a learning rate of 5e-3. An early stopping mechanism is adopted to prevent overfitting. 
Additionally, the speech LM conducts autoregressive generation throughout inference across all tasks. Beam search algorithm with a beam size of 5 is adopted during the sampling process.

\subsubsection{Pre-train, Fine-tune Paradigm}

In the ``pre-train, fine-tune'' paradigm, we train an expert downstream model for each task, utilizing the SSL speech representation as inputs. We adopt the same layer of the intermediate representation that derives the discrete units for the expert downstream model. ~\cite{chang2023exploring} indicates that using the SSL speech representation as input is a strong baseline compared to using discrete units as input.
Both HuBERT and mHuBERT adopt the HuBERT-base architecture~\cite{hsu2021hubert}, containing 12 Transformer layers with 95 million parameters.

For the expert downstream model's design, we follow SUPERB~\cite{DBLP:conf/interspeech/YangCCLLLLSCLHT21} and adjust the hidden dimension of the downstream model to achieve a lightweight expert model to compare with the prompting paradigm.
For speech classification tasks, we employ a linear model with a cross-entropy loss function.

We utilize a 2-layer LSTM with a hidden dimension of 256 for sequence generation tasks~\footnote{In the SUPERB setting, phoneme recognition utilizes CTC loss and a linear downstream model for frame-wise prediction. Following SUPERB, we also employ a linear model for the ``pre-train, fine-tune'' paradigm.}, and a 2-layer Transformer with an embedding size of 512 and 8 attention heads for speech generation tasks. For speech classification tasks, we adopted the Adam optimizer with a learning rate of 5e-3. For sequence generation tasks, we use the Adam optimizer with a learning rate of 1e-2 for phoneme recognition and 1e-4 for ASR. For speech generation tasks, we use the Adam optimizer with a learning rate of 5e-4 for 10 epochs.

\begin{table*}[!ht]
    \centering
    \caption{Performance comparison on sequence generation tasks for the ``pre-train, fine-tune paradigm'' (\textbf{FT}) and the ``prompting paradigm'' (\textbf{PT}). *: The learnable verbalizer introduces an extra small number of parameters, which is negligible here.
    }
    \renewcommand{\arraystretch}{0.8} 
    \resizebox{\linewidth}{!}{
    \begin{tabular}{cc|ccc|cc|ccc}
        \toprule
        \multicolumn{10}{c}{Sequence Generation Tasks} \\
        \midrule
        \midrule
        \multirow{2}{*}{Paradigm} & \multirow{2}{*}{Scenario}  & \multicolumn{3}{c|}{ASR-LibriSpeech}  & \multicolumn{2}{c|}{PR-LibriSpeech}  & \multicolumn{3}{c}{SF-AudioSnips} \\
        & & \#~Params. & WER~$\downarrow$ & CER~$\downarrow$ & \#~Params. & PER~$\downarrow$ &\#~Params. & CER~$\downarrow$ & F1~$\uparrow$\\
        \midrule
        FT & HuBERT~+~Expert & 2.9M &\textbf{15.67} & \textbf{4.55} & 2.6M &\textbf{5.34} & 2.9M & \textbf{40.08} & \textbf{78.53}\\
        PT & GSLM$_{\text{fixed}}$ & 4.5M & 34.17 & 26.14 & 4.5M & 21.10 & 4.5M & 66.90 & 59.47 \\
        \midrule
        \midrule
        FT & mHuBERT~+~Expert & 2.9M &14.44 & \textbf{4.43} & 2.6M & 12.42 & 2.9M & 32.24 & 85.26 \\
        \multirow{2}{*}{PT} & Unit mBART$_{\text{fixed}}$ & 2.6M & 13.85 & 5.91 & 2.6M & 5.16 & 2.6M & 33.09 & \textbf{87.20} \\
         & Unit mBART$_{\text{learn}}$ & 2.6M* & \textbf{11.56} & 5.13 & 2.6M* & \textbf{4.95} & 2.6M* &  \textbf{30.69} & 87.08 \\
        \bottomrule
    \end{tabular}
    }
    \vspace{-0.2cm}
\label{tab:sg_expresults}
\end{table*}

\begin{table*}[htbp]
    \centering
    \caption{Evaluation for the speech continuation task. Both \textnormal{GSLM} and \textnormal{Unit mBART} are in the prompting paradigm. Original: The ground truth corpus in the dataset. $r$: Conditional rate. Abbreviations: PPX (Perplexity), AB1 (Auto-BLEU-1), AB2 (Auto-BLEU-2), MOS (Mean Opinion Score), SIM (Speaker Similarity).}
    \resizebox{\linewidth}{!}{
    \begin{tabular}{c|ccccc|ccccc|ccccc}
    \toprule
    \multicolumn{16}{c}{Speech Generation Task (Speech Continuation)} \\
     \hline
     \hline
     \multirow{2}{*}{Scenario} & \multicolumn{5}{c|}{$r = 0.25$} & \multicolumn{5}{c|}{$r = 0.5$} & \multicolumn{5}{c}{$r = 0.75$}  \\
     & PPX~($\downarrow$) & AB1 & AB2 & MOS~($\uparrow$) & SIM~($\uparrow$) & PPX~($\downarrow$) & AB1 & AB2 & MOS~($\uparrow$) & SIM~($\uparrow$) & PPX~($\downarrow$) & AB1 & AB2 & MOS~($\uparrow$) & SIM~($\uparrow$) \\
    \hline
     GSLM  & \textbf{422.62} & 21.58 & 8.80 & 3.88 & 0.72 & \textbf{341.30} & 20.43 & 7.76 & 3.87 & 0.76 & \textbf{282.26} & 18.57 & 6.98 & 3.87 & 0.78 \\
    Unit mBART & 543.80 & 14.20 & 1.51 & \textbf{4.45} & \textbf{0.78} & 420.49 & 13.84 & 1.60 & \textbf{4.45} & \textbf{0.85} & 283.03 & 14.09 & 2.41 & \textbf{4.45} & \textbf{0.88} \\
    \hline
   Original & 202.92 & 13.9 & 2.42 & 4.47 & 0.86 & 202.92 & 13.9 & 2.42 & 4.47 & 0.95 & 202.92 & 13.9 & 2.42 & 4.47 & 0.98 \\
    \bottomrule
    \end{tabular}
    }
    \label{tab:speech_continuation}
\end{table*}

\begin{table}[t]
  \centering
  \caption{Evaluation on Spanish to English speech-to-speech translation.}
  \label{tab:speech_translation}
  \begin{tabular}{ccccc}
    \toprule
    \multicolumn{5}{c}{Speech Generation Task (Speech Translation)} \\
    \hline
    \hline
    Paradigm & Scenario & BLEU ($\uparrow$) & MOS ($\uparrow$) & \# Params.  \\
    \hline
    FT & mHuBERT + Expert & \multicolumn{3}{c}{$\times$} \\
    PT & GSLM & \multicolumn{3}{c}{$\times$} \\
    PT & Unit mBART & 15.89 & 4.32 & 10M \\
    \hline
    FT & Unit mBART & 18.47 & 4.33 & 353M \\
    \bottomrule
  \end{tabular}
\end{table}

\section{Results}
\subsection{Main Results}
\subsubsection{Speech Classification Tasks}
The comparison of the prompting paradigm (PT) and the ``pre-train, fine-tune'' paradigm (FT) for the speech classification tasks are shown in Table~\ref{tab:sc_full}. 
Our results indicate that the prompting method generally delivers competitive performance and often outperforms the fine-tuning approach. 
Specifically, for HuBERT and GSLM models, prompting outperforms fine-tuning in 6 out of 10 datasets (AR-SC, LT-SC, DM-SC, Grabo-SC, Fluent-SC, and Mustard++). 
For mHuBERT and mBART, prompting excels in 8 out of 10 datasets, including all datasets under SCR (with LT-SC achieving identical performance), IC, SD, and LID.

For the few tasks where prompting is slightly outperformed by fine-tuning in HuBERT and GSLM settings (Google-SC, and GFSound), the performance gap is minimal, within a 2\% difference. 
However, fine-tuning demonstrates a noticeable advantage in tasks like AcC and LID. 
This could be attributed to the loss of certain information, such as prosody, in the quantized HuBERT discrete units, leading to inferior GSLM performance compared to HuBERT with the downstream expert model.
In the mHuBERT and Unit mBART settings, while fine-tuning outperformed in 2 tasks, the performance difference in VAD is marginal (about 1.2\%). 

When assessing the effectiveness of utilizing a learnable verbalizer compared to a fixed one for prompting, it's observed that for GSLM, performance is enhanced in 6 out of 10 datasets (Google-SC, DM-SC, Grabo-SC, Mustard++, AccentDB, and GFSound). 
For unit mBART, the performances have improved or are on par in 7 datasets (Google-SC, AR-SC, LT-SC, DM-SC, Fluent-SC, Mustard++, and GFSound).

In summary, the prompting methods greatly match or exceed the performance of the fine-tuning approach across most speech classification tasks (6 outperform and 3 comparable for HuBERT and GSLM; 8 outperform and 1 comparable for mHuBERT and unit BART), except in accent classification. 

\subsubsection{Sequence Generation Tasks}
The experiment results of sequence generation tasks are shown in Table~\ref{tab:sg_expresults}. In sequence generation tasks, we observe that although prompting the decoder-only model GSLM can yield non-trivial results, it still underperforms compared to the fine-tuning paradigm by a substantial margin. 
The reasons are discussed in previous work~\cite{chang22e_interspeech}, including that quantizing speech into discrete units results in longer sequences, which might be difficult for a decoder model to handle.
In our preliminary study, even utilizing a learnable verbalizer for prompting GSLM does not show performance improvement.

On the other hand, surprisingly, prompting an encoder-decoder model like Unit mBART can achieve competitive performance, outperforming the fine-tuning paradigm in most scenarios, except for the metric CER in ASR, which falls behind by 0.7.
Furthermore, we can observe the effectiveness of introducing a learnable verbalizer in Unit mBART. 
For every metric other than the F1 score in Slot Filling (SF), there is a substantial improvement when comparing Unit mBART${_\text{learn}}$ with Unit mBART${_\text{fixed}}$. 
The analysis of the learnable verbalizer will be discussed in Section~\ref{sec:verb_ana}.
From GSLM to Unit mBART, the speech LM becomes better, and tasks that previously yielded poor results with GSLM can now yield favorable outcomes with Unit mBART. 
We anticipate that in the future, with more advanced speech LMs emerging, further performance improvement can be seen with the proposed prompting framework.

\subsubsection{Speech Generation Tasks}
In speech generation tasks, we focus on two tasks: Speech Translation (ST) and Speech Continuation (SC).
Our experiments show the effectiveness of prompting Unit mBART for speech translation, as detailed in Table~\ref{tab:speech_translation}. Speech-to-speech translation poses significant challenges, often requiring incorporating auxiliary tasks~\cite{jia19_interspeech, lee2022direct} or adopting an advanced speech LM~\cite{popuri22_interspeech}. Our results also support this observation: neither the prompting GSLM baseline nor the fine-tuning baseline with an expertly built mHuBERT model yielded reasonable results.
We experimented with various learning rates and downstream expert models; however, the fine-tuning baseline still yielded unsatisfactory outcomes.
On the other hand, prompting Unit mBART demonstrates proficiency in producing reasonable translations, as evidenced by its BLEU score. Compared to fine-tuning the whole Unit mBART, the prompting method has a performance drop but shows promising trainable parameter efficiency. Examples of the speech generation tasks can be found on the demo page~\footnote{Demo page: \url{https://ga642381.github.io/SpeechPrompt/speechgen}}.

Similarly, in speech continuation tasks, the SSL speech models (HuBERT and mHuBERT) paired with expert models did not produce reasonable results. 
As shown in Table~\ref{tab:speech_continuation}, for various conditional ratios $r$, we observed that prompting GSLM outperformed Unit mBART in terms of Perplexity (PPX), aligning with GSLM's pre-training for such tasks. 
Regarding the Auto-BLEU metric~\cite{lakhotia2021generative}, Unit mBART achieved scores comparable to the original utterances in the LJ Speech dataset. 
This suggests that the utterances generated by Unit mBART are as diverse as the oracle utterances, a challenge where GSLM falls behind.
Future research will explore varying the sampling temperature to enhance utterance generation quality, as discussed in~\cite{lakhotia2021generative}.
For speech quality, both GSLM and Unit mBART exhibit comparable MOS and speaker similarity to the original LJ Speech utterances~\footnote{Note that the quality of the generated speech is primarily controlled by the speech decoder, and high speaker similarity can be rooted in the fact that the pre-training data of the speech decoder includes LJ Speech.}. 

\begin{table*}[t]
    \centering
    \caption{Performance comparison on speech classification tasks with low resource dataset setting (10-shot) for the ``pre-train, fine-tune paradigm'' (\textbf{FT}) and the ``prompting paradigm'' (\textbf{PT}). The evaluation metric is accuracy.
    }
    \resizebox{\linewidth}{!}{
    \begin{tabular}{cc|ccccc|c|c|c|c|c}
        \toprule
        \multicolumn{12}{c}{Speech Classification Tasks (10-shot setting)} \\
        \midrule
        \midrule
        \multirow{2}{*}{Paradigm} & \multirow{2}{*}{Scenario} & \multicolumn{5}{c|}{SCR} & IC & SD & AcC & LID & VAD \\
        & & Google-SC & AR-SC & LT-SC & DM-SC & Grabo-SC  & Fluent-SC & Mustard++ & AccentDB & Voxforge & GFSound \\
        \midrule
        FT & HuBERT~+~Expert & 75.33 & 68.65 & \textbf{80.61} & 50.42 & 43.55 & 47.04 & 54.17 & \textbf{78.71} & 56.69 & \textbf{92.85} \\
        \multirow{2}{*}{PT} & GSLM$_{\text{fixed}}$ & \textbf{79.55} & 42.16 & 79.59 & \textbf{62.39} & 49.23 & \textbf{73.5} & 55.83 & 22.35 & 32.16 & 87.1 \\
         & GSLM$_{\text{learn}}$ & 77.15 & \textbf{90.00} & 68.36 & 61.50 & \textbf{86.50} & 72.87 & \textbf{65.83} & 26.71 & \textbf{85.41} & 89.30 \\
        \midrule
        \midrule
        FT & mHuBERT~+~Expert & 76.88 & 94.59 & 72.45 & 47.01 & 76.24 & 47.01 & 56.77 & 50.96 & 90.65 & \textbf{95.97} \\
        \multirow{2}{*}{PT} & Unit mBART$_{\text{fixed}}$ & \textbf{81.47} & \textbf{95.14} & 78.57 & \textbf{70.85} & \textbf{87.10} & 55.81 & \textbf{63.33} & 49.23 & \textbf{94.46} & 95.21 \\
         & Unit mBART$_{\text{learn}}$ & 80.46 & 93.51 & \textbf{86.74} & 64.96 & 85.85 & \textbf{64.90} & 53.33 & \textbf{57.32} & 93.83 & 92.84 \\
        \bottomrule
    \end{tabular}
    }
\label{tab:sc_few}
\end{table*}

\begin{figure*}[t]
\centering
\includegraphics[width=0.98\textwidth]{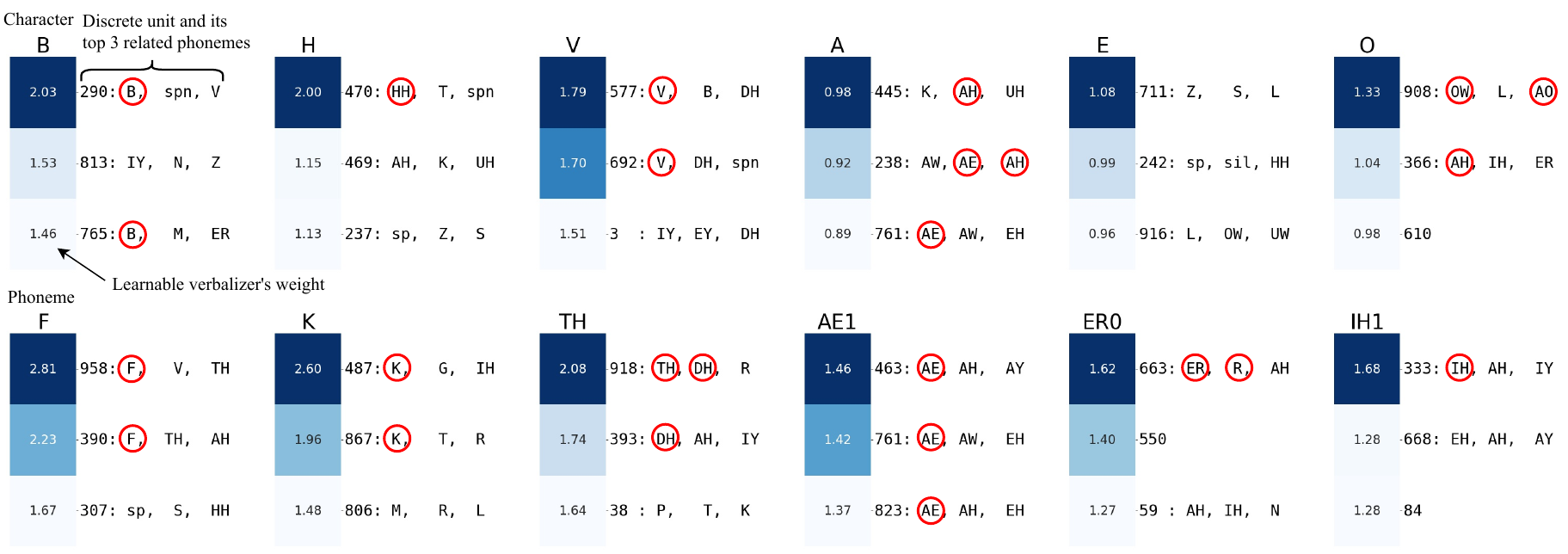}
\caption{Analysis of the Learnable Verbalizer. The top row presents heatmaps for the ASR task, with each subplot dedicated to the analysis of an individual character. The bottom row relates to the Phoneme Recognition (PR) task, with each subplot focused on a particular phoneme. 
The heatmaps show the weights that the learnable verbalizer assigns to the discrete units; for example, the phoneme ``AE1'' is most strongly linked to ``Unit 463''.
Besides the units, the top three phonemes with the highest correlation to the discrete units are listed, which is determined by forced alignment. 
This visualization illustrates the learnable verbalizer's ability to effectively utilize the information encoded in the discrete units to map to suitable labels. Related phonemes to the downstream tasks labels are circled.
}
\label{fig:char_phone_units}
\end{figure*}





\subsection{Few-shot Learning}
The prompting method has demonstrated its few-shot learning capabilities in the NLP field~\cite{DBLP:conf/eacl/SchickS21, liu2023pre} because of the inherent rich prior knowledge within language models. Similarly, speech LMs have already learned to comprehend discretized speech, that is, the discrete speech units.
This study extends the investigation into the few-shot learning abilities of the prompting method for speech LMs. 
We conduct 10-shot learning experiments. For both PT and FT, the trainable parameters are updated with the provided few-shot training data. Specifically, 10 samples per class were used as training data. The models are evaluated using the same testing set as the full dataset setting.

Table~\ref{tab:sc_few} illustrates the performance of the prompting method in comparison to the ``pre-train, fine-tune'' paradigm in a 10-shot learning scenario. 
The experiment shows that the prompting paradigm (PT) possesses robust few-shot learning capabilities and generally outperforms the fine-tuning paradigm (FT) in most speech classification tasks.
For HuBERT and GSLM, the FT method can only outperform the PT in 3 out of 10 tasks (LT-SC, AccentDB, GFSound). Meanwhile, for mHuBERT and Unit mBART, the FT method can only outperform the PT in 1 out of 10 tasks (GFSound).

Generally, in speech classification tasks under few-shot scenarios, prompting with Unit mBART achieves the best overall performance, showing top or near-top results in most tasks. Interestingly, we do not observe consistent performance improvement when utilizing a learnable verbalizer in these few-shot scenarios. We hypothesize that this might be due to the limited data, which causes challenges for the learnable verbalizer to extract the hidden information encapsulated in the discrete units effectively. The investigation of the underlying reason remains a future work.

\subsection{Verbalizer Analysis}
\label{sec:verb_ana}

In this study, we introduce an optional learnable verbalizer that bridges the gap between discrete units and the downstream tasks' labels. Prior research has shown that discrete units encapsulate acoustic and phonetic information~\cite{lakhotia2021generative, wells2022phonetic}. 
Thus, rather than employing a random mapping of the heuristic method in ASR and PR, it is more reasonable to employ a learnable verbalizer that discerns which discrete units correlate with specific labels, such as characters or phonemes.
The efficacy of the learnable verbalizer is presented in Fig.~\ref{fig:char_phone_units}. This figure demonstrates the capability of the learnable verbalizer in linking discrete units with characters for the ASR task, as displayed in the figure's first row, and with phonemes for the PR task, as illustrated in the second row.
The heatmaps display the weights $W$ from the learnable verbalizer in Euqation~\ref{eq:learn_verb}, with each map's right side indicating the connected discrete unit. 
Besides the units, the top three phonemes with the highest correlation to the discrete units are listed, as determined by forced alignment.
We observe that for a particular character, such as ``B,'' the verbalizer prefers discrete units with a strong association with the phoneme ``B.'' 
This pattern is consistent in the second row, which pertains to phoneme recognition tasks. 
Here, labels are connected to units with a high correspondence to the relevant phonemes.

\section{Discussion}
\label{sec:discussion}
We list observations, limitations, and future directions:

    \textbf{Architecture and Pre-training Task of Speech Language Models}: In the field of NLP, there is a growing trend towards employing decoder-only language models, particularly GPT variants, for a broad range of text generation tasks. However, based on the experimental results, we suggest that encoder-decoder models may offer distinct advantages for speech processing. We hypothesize that it is because many speech processing tasks require handling different modalities, especially the speech signal and text. The unique continuous characteristics of speech signals may be more effectively processed by an encoder. Therefore, encoder-decoder models are likely better suited for the first encoding of the speech signal into a compact representation, after which the decoder generates the desired output, whether it is a class label, text, or another speech signal. This observation aligns with recent work~\cite{chang2023prompting} comparing GSLM and Wav2Seq~\cite{wu2023wav2seq} models of similar sizes and datasets. The encoder-decoder model Wav2Seq demonstrates an advantage.

    However, it is important to note that the pre-training tasks of these models may also play a significant role in their performance. For instance, GSLM, which performs the next-token prediction task during pre-training, achieves promising results for the speech continuation task. Conversely, Unit mBART's pre-training task focuses on denoising, which may contribute to its superior performance in other speech processing tasks. The exploration of model architectures and their respective pre-training tasks remains an interesting and valuable direction for future research in the prompting paradigm.

    \textbf{Performance of Prompting Speech Language Models}: We have observed the competitive performance of the prompting Unit mBART model in both speech classification and generation tasks. Notably, in speech generation tasks, relying solely on an SSL speech model does not yield satisfactory performance. However, a discernible performance gap still exists between the prompting and fine-tuning paradigms, especially the sequence generation task. Taking SUPERB as an example, the setting involves performing a weighted sum over the representations of each layer of SSL speech models and building an expert model on top of this, along with adopting a customized loss for the downstream task. Although such a setting requires considerable human labor and computational resources, its performance is competitive. In Table.~\ref{tab:superb}, we list the ranking of the prompted Unit mBART for the ASR task and compare it with the SSL speech models on SUPERB.
    
\begin{table}[t]
    \centering
    \caption{Sequence generation task performance. The models are ordered based on the ASR performance.}
    \resizebox{\linewidth}{!}{
    \begin{tabular}{lcccc}
        \toprule
        \textbf{Model} & {\textbf{ASR (WER~$\downarrow$)}} & {\textbf{SF (CER~$\downarrow$)}} & {\textbf{SF (F1~$\uparrow$)}} & \#~params\\
        \midrule
        FBANK & 23.18 & 52.94 & 69.64 & 43M\\
        modified CPC & 20.18 & 49.91 & 71.19 & 43M\\
        TERA & 18.17 & 54.17 & 67.5 & 43M\\
        vq-wav2vec & 17.71 & 41.54 & 77.68 & 43M\\
        wav2vec & 15.86 & 43.71 & 76.37 & 43M\\
        DeCoAR 2.0 & 13.02 & 34.73 & 83.28 & 43M\\
        \rowcolor{gray!25} \textbf{Unit mBART$_{\text{learn}}$} & 11.56 & 30.69 & 87.08 & \textbf{2.89M}\\
        wav2vec 2.0 Base & 6.43 & 24.77 & 88.3 & 43M\\
        HuBERT Base & 6.42 & 25.2 & 88.53 & 43M\\
        WavLM Base & 6.21 & 22.86 & 89.38 & 43M\\
        data2vec Large & \textbf{3.36} & \textbf{22.16} & \textbf{90.98} & 43M\\
        \bottomrule
    \end{tabular}
    }
    \vspace{-0.4cm}
    \label{tab:superb}
\end{table}

    \textbf{Develop Advanced Speech Language models}: Speech language models are currently in their nascent stage of development compared to text-based language models. The proposed prompt framework, although effective in motivating speech LMs, may not achieve exceptional performance. However, with advancements in speech LMs, such as the transition from GSLM to Unit mBART, there has been a significant improvement in prompt performance. Particularly, tasks that were previously challenging for GSLM now exhibit improved performance with Unit mBART. We anticipate the emergence of even more promising speech LMs in the future.

    Moreover, this paper primarily focuses on textless speech LMs, where the model adapts to various tasks through prompt optimization. Achieving true 0-shot inference remains a challenging but compelling goal within the field of speech processing. Recent advances, such as instruction-tuned speech LMs~\cite{huang2024dynamic, gong2023joint, tang2024salmonn}, although they might be restricted to a specific language with written text, highlight promising avenues toward achieving 0-shot adaptation by guiding the LMs with text instructions.

    \textbf{Beyond Content Information}: Current speech LMs do not fully capture speaker and emotion information, posing a challenge for tasks beyond content-related aspects. In scenarios where preserving speaker and emotion information is possible, we plan to explore the integration of plug-and-play modules specifically designed to incorporate speaker and emotion details into the framework. Looking ahead, we anticipate that future speech LMs will incorporate and leverage these additional factors and better handle speaker and emotion-related aspects in speech generation tasks. Google's latest speech LM~\cite{nachmani2023lms} tries to include such information.

\vspace{-0.2cm}

\section{Conclusion}
\label{sec:conclusion}
In this paper, we investigate how prompting can leverage the generative capabilities of speech language models (speech LMs) for solving a wide range of speech processing tasks. 
Our approach includes minimal trainable parameters to guide the speech LMs within a unified framework, achieving competitive performance compared to the fine-tuning paradigm while keeping the benefits of the prompting paradigm.
The proposed framework exhibits several desirable characteristics, including its textless nature, versatility, efficiency, transferability, and affordability. 
To demonstrate our framework's capabilities, we study the decoder-only GSLM and encoder-decoder Unit mBART as case studies. 
We conduct experiments on three distinct types of speech processing tasks: speech classification, sequence generation, and speech generation.
Also, the proposed framework shows promising results in the few-shot scenario.
We observe a trend that as more advanced speech LMs are developed, the performance of prompting will significantly improve.
We also discuss the limitations and future directions of prompting speech LMs. 
With the imminent arrival of advanced speech LMs, our unified framework holds immense potential in terms of efficiency and effectiveness, standing on the shoulders of giants.

\vspace{-0.2cm}
\bibliographystyle{IEEEtran}
\bibliography{refs}

\end{document}